\documentclass{article}

\usepackage{arxiv}

\usepackage[utf8]{inputenc} % allow utf-8 input
\usepackage[T1]{fontenc}    % use 8-bit T1 fonts
\usepackage{hyperref}       % hyperlinks
\usepackage{url}            % simple URL typesetting
\usepackage{booktabs}       % professional-quality tables
\usepackage{amsfonts}       % blackboard math symbols
\usepackage{microtype}      % microtypography
\usepackage{lipsum}
\usepackage{amssymb}
\usepackage{amsmath,amsthm}
\usepackage{graphicx}
\usepackage{enumerate}
\usepackage{subfigure}

\graphicspath{
    {./fig/}{./fig2/}{./fig-all/}
}

\title{Geodesic Centroidal Voronoi Tessellations:% on Manifold Meshes:
\\Theories, Algorithms and Applications}

\author{
  Zipeng Ye, Ran Yi, Minjing Yu, Yong-Jin Liu\thanks{Corresponding author}\\
  Department of Computer Science and Technology,\\
  Tsinghua University, China\\
   \And
 Ying He \\
  School of Computer Engineering,\\
  Nanyang Technological University, Singapore\\
}

\begin{document}
\maketitle

\begin{abstract}
Nowadays, big data of digital media (including images, videos and 3D graphical models) are frequently modeled as low-dimensional manifold meshes embedded in a high-dimensional feature space.
In this paper, we summarized our recent work on geodesic centroidal Voronoi tessellations (GCVTs), which are intrinsic geometric structures on manifold meshes. We show that GCVT can find a widely range of interesting applications in computer vision and graphics, due to the efficiency of search, location and indexing inherent in these intrinsic geometric structures. Then we present the challenging issues of how to build the combinatorial structures of GCVTs and establish their time and space complexities, including both theoretical and algorithmic results.
\end{abstract}

% keywords can be removed

\section{Introduction}

Let $(X,d)$ be a metric space, where $X$ is a point set and $d:X\times X\rightarrow\mathbb{R}$ is a metric.
Given an open subset $\Omega \subseteq X$, a set $\{ V_i \}_{i=1}^k$ is called a \emph{tessellation} of $\Omega$ if $V_i \cap V_j = \emptyset$ for $i \neq j$ and $\cup_{i=1}^k \overline{V_i} = \overline{\Omega}$, where $\overline{A}$ is the closure of $A$.
Given a set of points $\{ g_i \}_{i=1}^k$ in $\Omega$, the \emph{Voronoi cell} corresponding to the point $g_i$ is defined as
\begin{equation}
\hat{V_i} = \{ x \in \Omega\; \big|\; d(x, g_i) < d(x, g_j) \;\mbox{for}\,j=1,\dots,k,\,j \neq i \}.
\end{equation}
Elements of $\{ g_i \}_{i=1}^k$ are called \emph{generators}.

Since 1644 (Part III of Principia Phiolosophiae, written by Descartes), Voronoi tessellations had been well studied in the Euclidean space $\mathbb{R}^n$, $n\in\mathbb{Z}^+$ \cite{OBS2000}, in which Voronoi tessellations (as well as their dual structures, well known as Delaunay triangulations) had played a central role as fundamental geometric structures.
Voronoi tessellations had also been studied in spaces with non-Euclidean metrics, including spheres \cite{Augenbaum1985}, hyperbolic spaces \cite{Onishi1996} and the general Riemannian manifolds \cite{Leibon2000}. Recently, due to the flourishes of big media data from digital sampling, more and more data are appearing in the form of manifold meshes \cite{Seung2000}.
Quite different from parametrized 2-manifolds or general Riemannian manifolds that are generally $C^\infty$ smooth (or at least $C^2$ smooth) \cite{Boothby1975}, manifold meshes are only $C^0$.
In our study, we adopt the discrete geodesic metric \cite{Mitchell87}.

In this paper, we summarize our recent work on geodesic centroidal Voronoi tessellations (GCVT) --- which are provable uniform tessellations on manifold meshes --- and we show that they can be used to generate uniform remeshing in computer graphics and build content-sensitive superpixels/supervoxels for images and video in computer vision applications.

Before we introduce GCVTs, we present two close concepts of CVT and RCVT in related work.

\section{Related Work}

\subsection{Centroidal Voronoi Tessellations (CVT)}

CVT had been well studied in science and engineering, with a wide range of applications including data compression in digital image processing, optimal quadrature in numerical methods, quantization and clustering in machine learning, finite difference methods in solid mechanics and fluid dynamics, distribution of resources in operational research, cellular patterns in biology, and the territorial behavior of animals; see \cite{Du1999Centroidal} for an excellent survey.

Let V be a finite region in $\mathbb{R}^n$. The mass centroid $m$ of $V$ is defined by
\begin{equation}
m \triangleq \frac{\int_{x \in V} x\rho(x)dx}{\int_{x \in V} \rho(x)dx},
\label{eq:mass_centroid}
\end{equation}
where $\rho$ is a density function defined in $V$.
Given $k$ points $\{ g_i \}_{i=1}^k$ in a domain $\Omega\subseteq\mathbb{R}^n$, we can define the Voronoi region $\hat{V_i}$ corresponding to each $g_i$ based on the Euclidean metric $d_E$ and the Voronoi tessellation $\{ \hat{V_i} \}_{i=1}^k$ of $\Omega$. Let $m_i$ be the mass centroid of each Voronoi region $\hat{V_i}$. A Voronoi tessellation is called CVT if all the generators are mass centroids, i.e.
\begin{equation}
g_i = m_i,\ i=1,\dots,k.
\label{eq:cvt_condition}
\end{equation}
Arbitrarily chosen generators are usually not the mass centroids of their associated Voronoi regions so an arbitrary Voronoi tessellation cannot be a CVT.
It can be shown that CVT minimizes the following energy (a.k.a. CVT energy functional):
\begin{equation}
\varepsilon_E(\{ (p_i, V_i) \}_{i=1}^k) = \sum_{i=1}^k \int_{x \in V_i} \rho(x)d_E^2(x, p_i)dx,
\label{eq:CVT_energy}
\end{equation}
where $\rho$ is a density function defined in $\Omega$, $\{ V_i \}_{i=1}^k$ is an arbitrary tessellation and $\{ p_i \}_{i=1}^k$ is any set of $k$ points in $\Omega$.
To compute CVT, various local methods had been surveyed in \cite{Du1999Centroidal}. In particular, the Lloyd method \cite{Lloyd} is a simple yet effective local method that iteratively computes mass centroids and Voronoi tessellation.

\subsection{Constrained CVT for Surfaces}

To extend the domain partitioning of CVT from Euclidean space to general spaces, Du et al. \cite{Du2003} proposed the constrained CVT (CCVT) that works on a compact and continuous surface $S\subset\mathbb{R}^N$ defined by
\begin{equation}
S\triangleq\{x\in\mathbb{R}^N:g_0(x)=0\ \mbox{and}\ g_i(x)\leq 0,\ \mbox{for}\ i=1,2,\cdots,m\}
\label{eq:CCVT_surf}
\end{equation}
where $\{g_i\}_{i=0}^m$ are some continuous functions.

Given a set of points $\{ g_i \}_{i=1}^k\in S$, the constrained Voronoi region $\hat{V_i}$ corresponding to each $g_i$ based on the Euclidean metric $d_E$ and restricted in $S$ is defined as
\begin{equation}
\hat{V_i}\triangleq\{ x \in S\ \big|\; d_E(x, g_i) < d_E(x, g_j) \;\mbox{for}\ j=1,\dots,k,\,j \neq i \}.
\label{eq:CCVT_vcell}
\end{equation}
The constrained mass centroid $m_i$ of $\hat{V_i}$ on $S$ is defined to be the solution of the following problem:
\begin{equation}
\min_{x\in S}\varepsilon_i(z),\quad \mbox{where}\ \varepsilon_i(z)=\int_{x \in \hat{V_i}}\rho(y)d_E^2(x,z)dx
\label{eq:CCVT_centroid}
\end{equation}
A tessellation on $S$ is called CCVT if and only if the set $\{ g_i \}_{i=1}^k\in S$ are both the generators and constrained mass centroids of the tessellation $\{ \hat{V_i} \}_{i=1}^k$.

The applications of CCVT including polynomial interpolation and numerical integration on the sphere are illustrated in \cite{Du2003}.

\begin{figure}
    \centering
    \includegraphics[width=\textwidth]{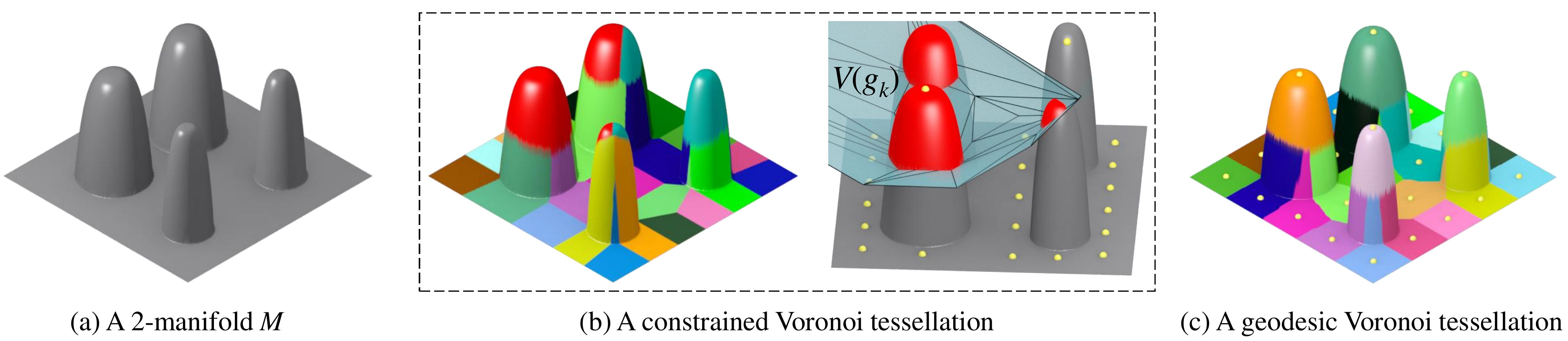}
    \caption{(b) shows a constrained Voronoi tessellation of 30 point generators on the 2-manifold (a) and its Voronoi cell $V(g_k)$ is disconnected; i.e., it consists of three disjoint components (shown in red areas). Given the same set of point generators, each cell in geodesic Voronoi tessellation (c) is guaranteed to be connected.}
    \label{fig:CCVT-GCVT}
\end{figure}

\begin{figure}
    \centering
    \includegraphics[width=0.3\textwidth]{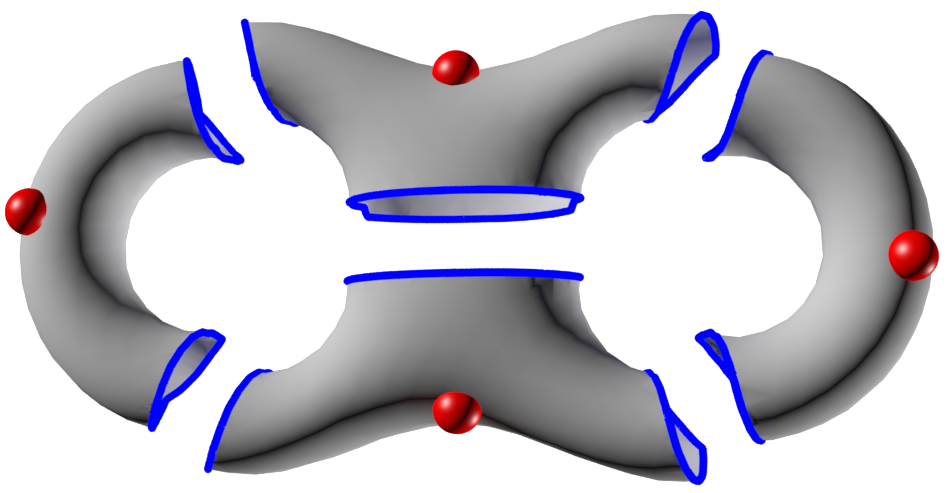}
    \caption{A geodesic Voronoi tessellation on a genus-2 manifold whose cells are multiply connected.}
    \label{fig:multi-connected}
\end{figure}

\section{Geodesic Centroidal Voronoi Tessellations (GCVT)}

Given a k-dimensional compact differentiable manifold $M$, a set $\{ V_i \}_{i=1}^k$ is called a \emph{tessellation} of $M$ if $V_i \cap V_j = \emptyset$ for $i \neq j$ and $\cup_{i=1}^k \overline{V_i} = \overline{M}$. Denote by $d_g(a,b)$ the geodesic distance between $a$ and $b$ in $M$. Given a set of points $\{ g_i \}_{i=1}^k$ in $M$, the \emph{geodesic Voronoi region} corresponding to the point $g_i$ is defined as
\begin{equation}
\hat{V_i} = \{ x \in M\; \big|\; d_g(x, g_i) < d_g(x, g_j) \;\mbox{for}\,j=1,\dots,k,\,j \neq i \}.
\end{equation}
Elements of $\{ g_i \}_{i=1}^k$ are called \emph{generators}. The set of geodesic Voronoi regions $\{ \hat{V_i} \}_{i=1}^k$ is called geodesic Voronoi tessellation of $M$. A geodesic Voronoi region is a connected domain and is a non-empty compact set \cite{Yong2011Construction}. For each geodesic Voronoi region $\hat{V_i}$, the \emph{nominal mass centroid} $m_i$ of $\hat{V_i}$ on $M$ is defined to be the solution of the following problem
\begin{equation}
\label{eq:nominal_mass_centroid}
\mathop{\min}_{z \in M} \varepsilon_i(z), \mbox{where}\; \varepsilon_i(z) = \int_{x \in \hat{V_i}}\rho(x)d_g(x, z)dx.
\end{equation}
It can be shown that $\varepsilon_i(z)$ is continuous and the domain $M$ is compact so $\varepsilon_i(z)$ has at least one global minimum in $M$. Therefore, the nominal mass centroid exists.

Given $k$ points $\{ g_i \}_{i=1}^k$, we can define the geodesic Voronoi region $\hat{V_i}$ corresponding to each $g_i$ and geodesic Voronoi tessellation $\{ \hat{V_i} \}_{i=1}^k$ on $M$.
For each geodesic Voronoi region $\hat{V_i}$, its nominal mass centroid $m_i$ is defined by Eq.\eqref{eq:nominal_mass_centroid}. We call the geodesic Voronoi tessellation as \emph{geodesic centroidal Voronoi tessellation} (GCVT) if all the generators are nominal mass centroids, i.e.
\begin{equation}
\label{eq:gcvt_condition}
g_i = m_i,\; i=1,\dots,k.
\end{equation}

We define an energy function $\varepsilon_g$ from any tessellation $\{ V_i \}_{i=1}^k$ of $M$ and $k$ points $\{ p_i \}_{i=1}^k\in M$:
\begin{equation}
\label{eq:gcvt_energy}
\varepsilon_g(\{ (g_i, V_i) \}_{i=1}^k) = \sum_{i=1}^k \int_{x \in V_i} \rho(x)d_g^2(x, g_i)dx.
\end{equation}
We call $\varepsilon_g$ the GCVT \emph{energy functional}. It can be shown that the necessary condition for $\varepsilon_g$ being minimized is that $\{ (g_i, \hat{V_i}) \}_{i=1}^k$ is a GCVT \cite{Liu2017Intrinsic} so we can obtain a GCVT by optimizing the GCVT energy functional. The theoretical results for the combinatorial structures of geodesic Voronoi tessellations will be presented in Section \ref{sec:theory} and the algorithm for finding a GCVT will be introduced in Section~\ref{sec:algorithms}.

\begin{figure}
\centering
\subfigure[Stretching map]{\includegraphics[width=.32\textwidth]{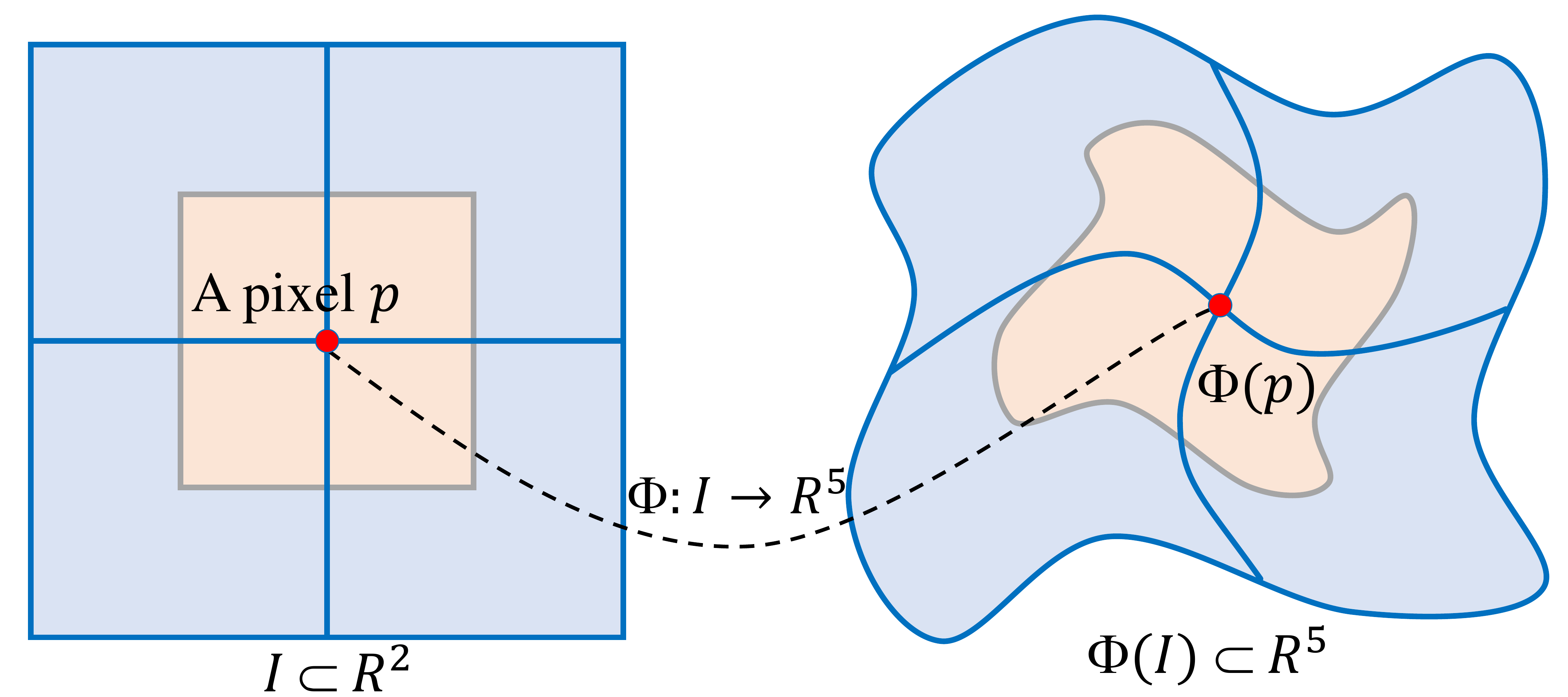}}\hspace{12pt}%
\subfigure[Content-sensitive superpixels via image manifold]{\includegraphics[width=.65\textwidth]{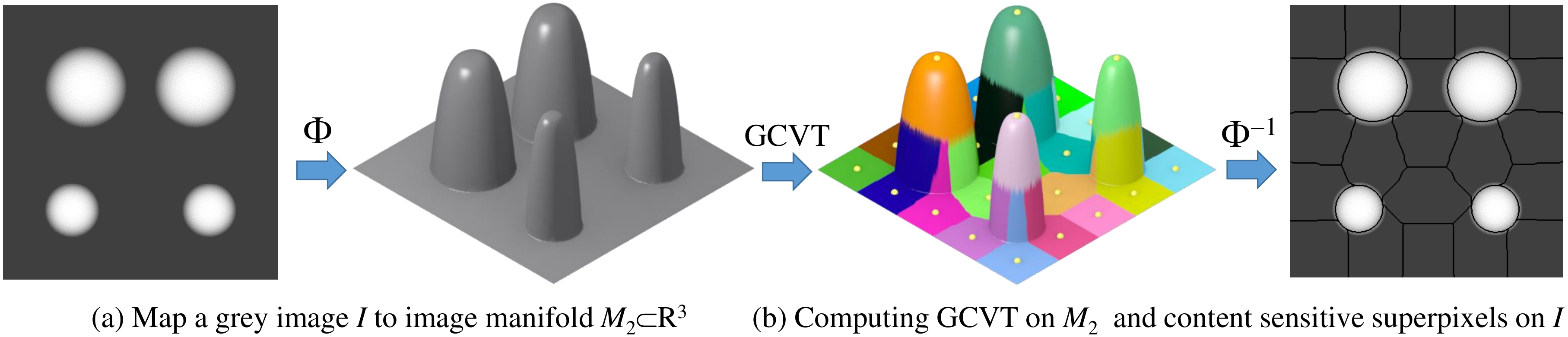}}%
\caption{(a) The map $\Phi$ that stretches an image $I\subset\mathbb{R}^2$ to an image manifold $M_2\subset\mathbb{R}^5$.
(b) A uniform tessellation on $M_2$ such as GCVT naturally induces content-sensitive superpixels in $I$ via $\Phi^-1$; for an easy illustration, a gray image is used for a mapping in $\mathbb{R}^3$.}
\label{fig:image-manifold}
\end{figure}

\subsection{Comparison of CVT, CCVT and GCVT}

At the first glance, CVT, CCVT and GCVT are all uniform tessellations with similar formulations and we call the Voronoi regions of any tessellation as \emph{cells}. CVT is defined in Euclidean spaces and its cells are convex polygon/polyhedra that are the intersection of half spaces. Both CCVT and GCVT are defined on manifold domains. However, CCVT needs that the manifold is embedded in a Euclidean space such that the Euclidean metric $d_E$ can be used in Eq.(\ref{eq:CCVT_vcell}). Then we say CCVT is {\it extrinsic}, which could also be interpreted as the intersection between the manifold and the Voronoi tessellation in Euclidean space. Therefore, the cells in CCVT may be disconnected or multiply-connected\footnote{A region is {\it simply connected} if any simple closed curve in it can be continuously shrunk into a point without leaving the region. A connected region that is not simply connected is called {\it multiply connected}.}; see Figures \ref{fig:CCVT-GCVT} and \ref{fig:multi-connected} for an illustration.

GCVT is defined on a compact differentiable manifold without referring to an embedded Euclidean space. Since GCVT only relies on the geodesic metric, it is {\it intrinsic}. Compared to CCVT, all the cells in GCVT are guaranteed to be connected.

\section{Applications}

Before we present theoretic and algorithmic results of GCVT, we state three representative applications, showing that GCVT is a useful tool in computer vision and graphics.

\subsection{Content-Sensitive Superpixels}

Image pixels are only the units of image capturing device, but not optimized for image content presentation.
Superpixels are a dense over-segmentation of image, which capture well image features and can serve as perceptually meaningful atomic regions for images.
Superpixels can be used as a preprocessing for reducing the complexity of subsequent image processing tasks, which includes segmentation\cite{IMSLIC1_Yu2011Entropy}, contour closure\cite{IMSLIC2_Levinshtein2010Optimal}, object location \cite{IMSLIC3_Fulkerson2009Class}, object tracking \cite{IMSLIC4_Shu2011Superpixel}, stereo 3D reconstruction \cite{IMSLIC5_Mi2010Multi}, and many others. See \cite{StutzHL18cviu} for a comprehensive survey.

To serve as perceptually meaningful atomic regions, superpixels generally have the following characteristics \cite{Liu2017Intrinsic}:
\begin{enumerate}[(1)]
\item \emph{Partition}: each pixel in the image is assigned to exactly one superpixel so superpixels are a partition of the image;
\item \emph{Connectivity}: each region of superpixel is simply connected;
\item \emph{Compactness}: in the non-feature region, superpixels are regular in shape and uniform in size;
\item \emph{Feature preservation}: superpixels should adhere well to image boundaries for preserving feature;
\item \emph{Content sensitivity}: the density of superpixels is adaptive to the variety of image contents;
\item \emph{Performance}: superpixels should be computed in a low cost of time and space.
\end{enumerate}

In \cite{Liu2016cvpr,Liu2017Intrinsic}, we propose an image manifold that maps a color image $I$ from $\mathbb{R}^2$ to a 2-manifold $M_2$ embedded in the 5-dimensional combined image and colour space $\mathbb{R}^5$:
\begin{equation}
\Phi(u, v) = (\lambda_1 p, \lambda_2 \mathbf{c}) = (\lambda_1 u, \lambda_1 v, \lambda_2 l, \lambda_2 a, \lambda_2 b),
\end{equation}
where $I(u,v)$ is a color image with pixel positions $p=(u,v)$, $\mathbf{c}(p)=(l(u, v), a(u, v), b(u, v))$ is the color at the pixel $p$ in CIELAB color space, $\lambda_1$ and $\lambda_2$ are global stretching factors.
The area elements in the image manifold $M_2$ are a good measure of the content density in the image $I$. Then a uniform tessellation such as GCVT on $M_2$ naturally induce good content-sensitive superpixels in $I$. See Figure \ref{fig:image-manifold} for an illustration.

\begin{figure}
    \centering
    \includegraphics[width=0.6\textwidth]{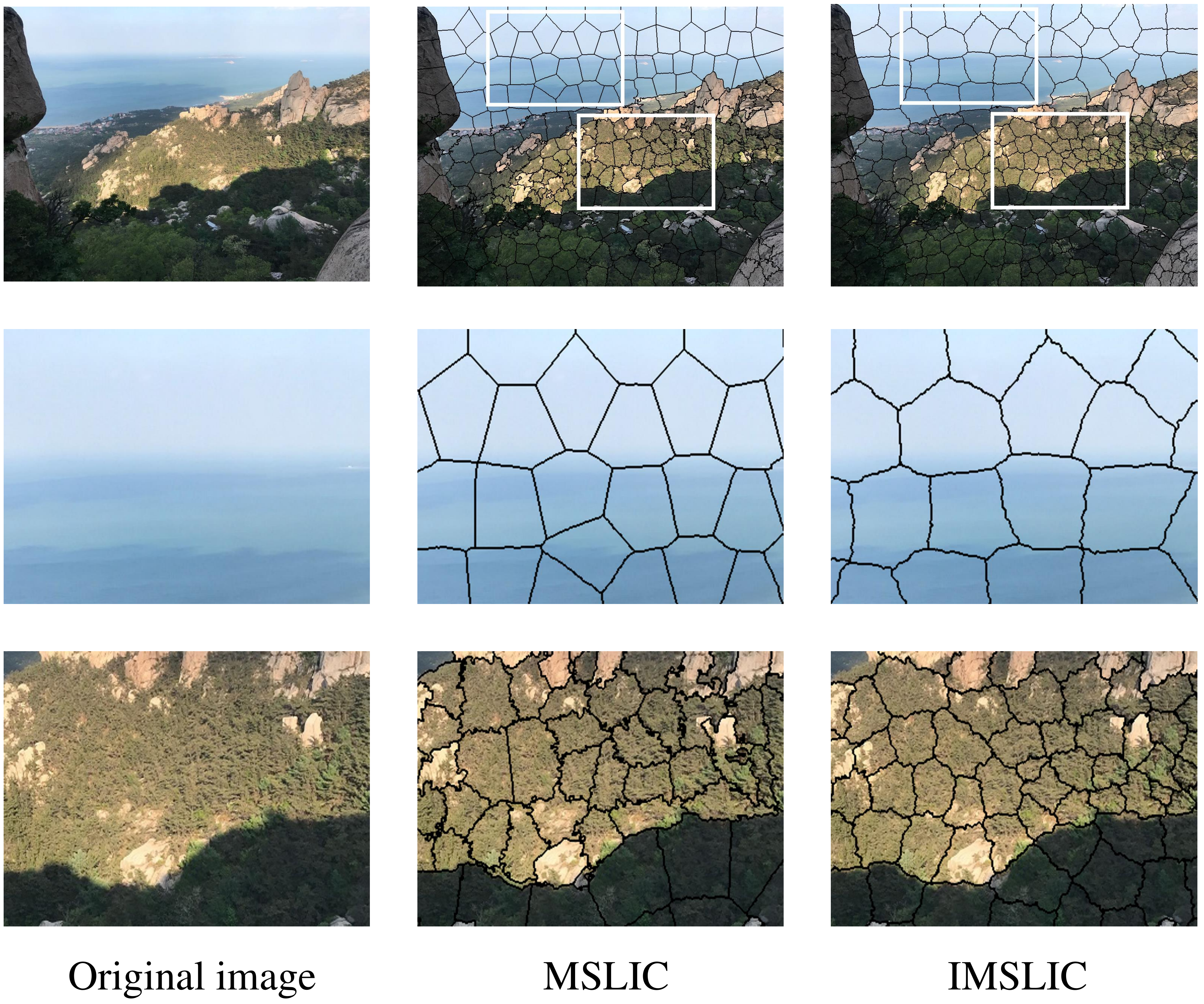}
    \caption{Some examples of two content-sensitive superpixels (MSLIC \cite{Liu2016cvpr} and IMSLIC \cite{Liu2017Intrinsic}) based on the uniform tessellation on the image manifold.}
    \label{fig:superpixels}
\end{figure}

GCVT is a powerful tool for superpixels due to the following reasons:
\begin{enumerate}[(1)]
\item \emph{Partition}: GCVT is a tessellation of $M_2$ (and also $I$ due to the one-to-one mapping $\Phi$);
\item \emph{Connectivity}: each cell in GCVT is guaranteed to be connected;
\item \emph{Compactness}: cells in GCVT are regular and uniform in non-feature regions;
\item \emph{Feature preservation}: the feature regions (such as object boundary) have a large color variation and therefore lead to a large stretching/area on $M_2$. The larger the area in $M_2$, the higher possibility that a cell boundary passes through it;
\item \emph{Content sensitivity}: content-dense regions have high intensity or color variation, and then larger area on $M_2$. Given uniform tessellation on $M_2$, the superpixels will be smaller in content-dense regions. Similarly, content-sparse regions have large superpixels;
\item \emph{Performance}: we propose efficient computation methods in Section x that can quickly approximate GCVTs.
\end{enumerate}
See Figure \ref{fig:superpixels} for some qualitative results of content-sensitive superpixels.

\begin{figure}
    \centering
    \includegraphics[width=.6\textwidth]{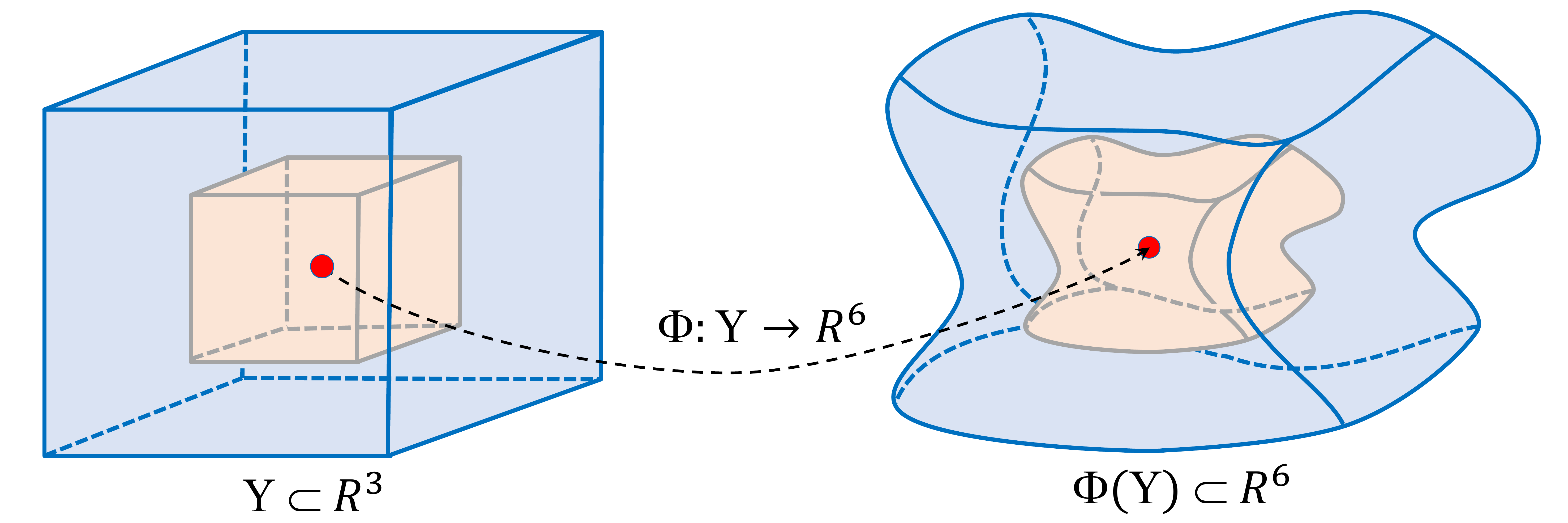}
    \caption{The stretching map $\Phi:\Upsilon\rightarrow M_3\subset\mathbb{R}^6$ maps a video $\Upsilon\subset\mathbb{R}^3$ into a 3-manifold $M_3\subset\mathbb{R}^6$.}
    \label{fig:mapping3d}
\end{figure}

\subsection{Content-Sensitive Supervoxels}

Akin to superpixels for images, Supervoxels are perceptually meaningful atomic regions in videos, obtained by grouping similar voxels that exhibit coherence in both appearance and motion.
Superpixels over-segment a video in the spatiotemporal domain while well preserving its structural content.

To compute content-sensitive supvoxels, Yi et al. \cite{Yi2018cvpr} extend the image manifold concept to the video manifold using the stretching map $\Phi:\Upsilon\rightarrow M_3\subset\mathbb{R}^6$ (Figure \ref{fig:mapping3d}):
\begin{equation}
\Phi(\Upsilon) = \Phi(u, v, t) = (\lambda_1 u, \lambda_1 v, \lambda_2 t,  \lambda_3 \mathbf{c}),
\end{equation}
where the 3-manifold $M_3$ is embedded in the 6-dimensional combined video and colour space $\mathbb{R}^6$, $\Upsilon$ is a video with $N$ voxels, $\upsilon(u,v,t)\in\Upsilon$ is a voxel with frame index $t$ and the pixel position $(u,v)$ in the frame, $\mathbf{c}(\upsilon)=(l(u,v,t), a(u,v,t), b(u,v,t))$ is the color of $\upsilon(u,v,t)$ in CIELAB color space $\lambda_1$, $\lambda_2$ and $\lambda_3$ are global stretching factors.

\begin{figure}
    \centering
    \includegraphics[width=0.4\textwidth]{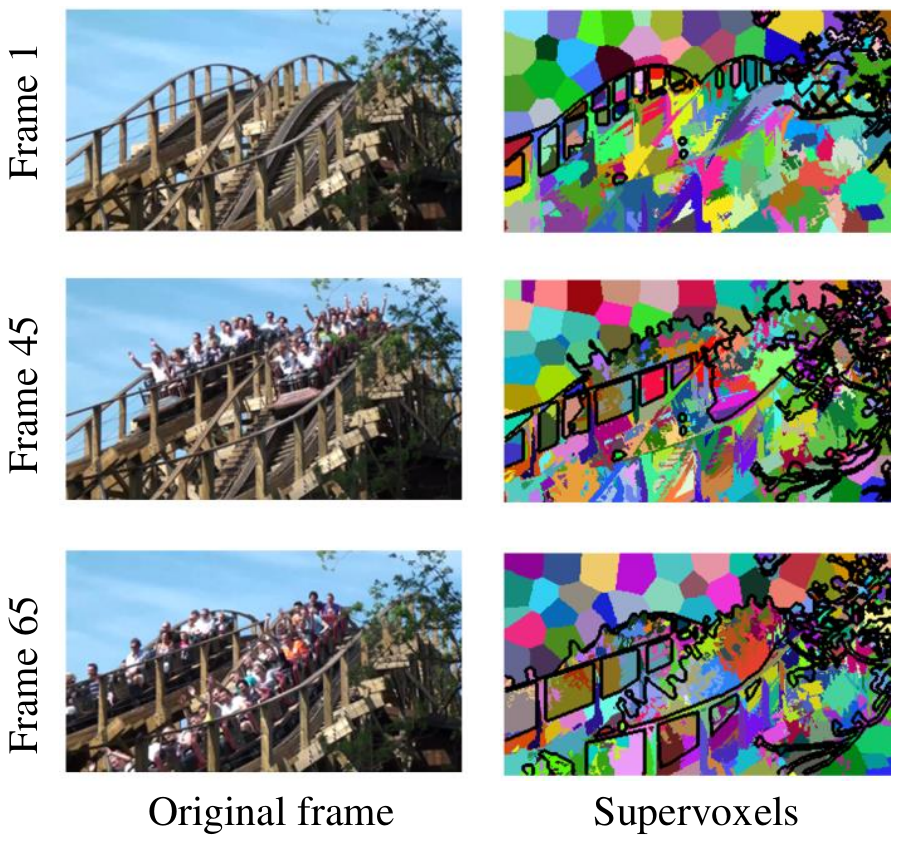}
    \caption{Examples of content-sensitive supervoxels \cite{Yi2018cvpr} based on the uniform tessellation on the video manifold. For an easy illustration, supervoxels are clipped in each frame and shown as cross-sectional superpixels.}
    \label{fig:supervoxels}
\end{figure}

At the place where the color variation is large in $\Upsilon$, $\Phi$ maps a unit voxel into a large volume in $M_3$. Therefore, the volume elements in $M_3$ offer a good measure of the content density in $\Upsilon$. In a similar way to content-sensitive superpixels, a uniform tessellation such as GCVT on $M_3$ will naturally induce content-sensitive supervoxels in $\Upsilon$, i.e., supervoxels are typically larger and longer in content-sparse regions (i.e., with homogeneous appearance and motion), and smaller and shorter in content-dense regions (i.e., with high variation of appearance and/or motion). See Figure \ref{fig:supervoxels} for an illustration.

\begin{figure}
\centering
\subfigure[GCVT]{\includegraphics[width=.25\textwidth]{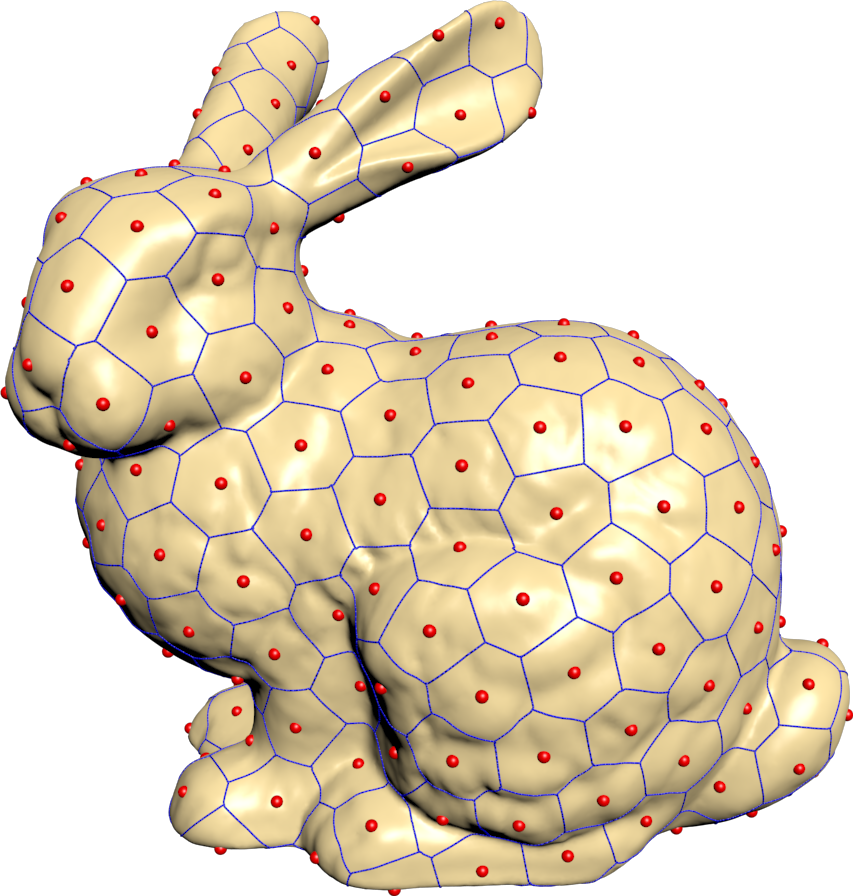}}\hspace{12pt}%
\subfigure[Dual IDT]{\includegraphics[width=.25\textwidth]{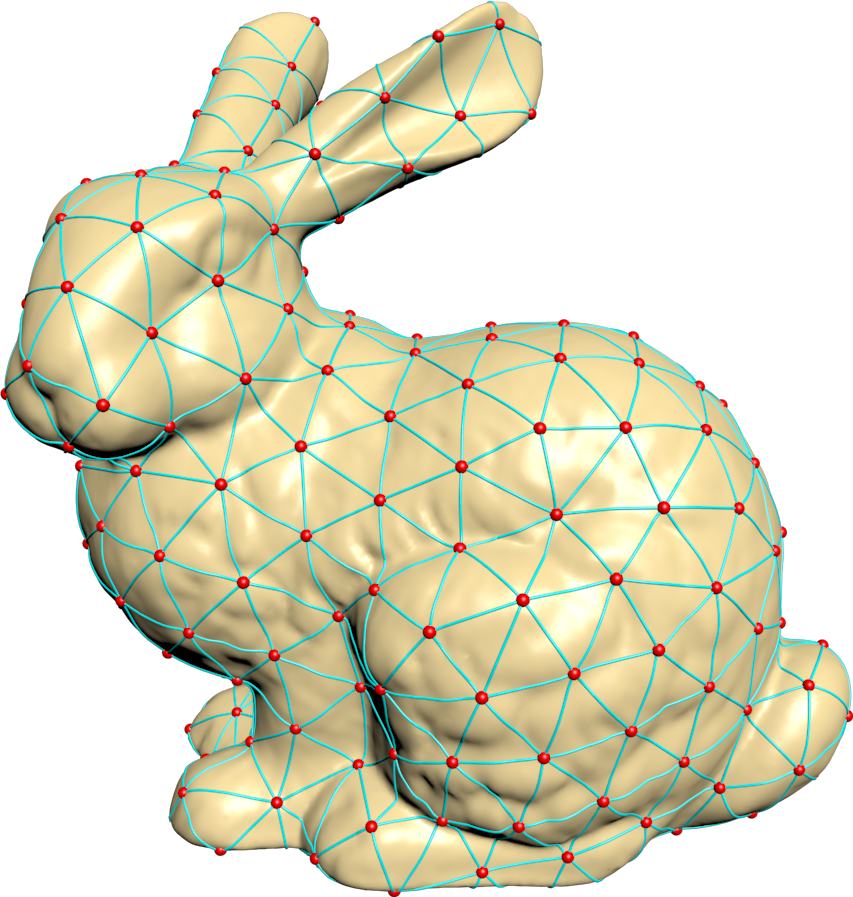}}%
\caption{(a) GCVT with a small number of generators \cite{Liu2016siga}. (b) Its dual intrinsic Delaunay triangulation (IDT) \cite{Liu2017CID}.}
\label{fig:remesh}
\end{figure}

\subsection{Low-Resolution Remeshing}

In computer graphics, 3D shapes are usually represented by triangular 2-manifold. In many engineering applications such as finite element analysis, a high quality mesh with almost congruent triangles is desired. To convert an arbitrary triangular mesh into a high quality mesh while preserving geometric shapes, remeshing techniques are developed. Low-resolution remeshing is to generate a mesh with a small number of vertices and the vertex size approaching the feature size of the original high-resolution mesh. Due to the existence of thin-shell structures, Voronoi tessellations based on Euclidean metric frequently results in disjoint fragments in a Voronoi cell. See Figure \ref{fig:CCVT-GCVT}b for an illustration.

GCVT can guarantee that each Voronoi cell is connected and thus is suitable for this low-resolution remeshing task. A globally optimized GCVT \cite{Liu2016siga} can generate uniform tessellations with regular cells on a given high-resolution mesh. We propose an efficient sampling criterion such that the intrinsic Delaunay triangulation due to the GVT exists \cite{Liu2017CID}. Therefore, it provides an efficient solution to low-resolution remeshing. See Figure \ref{fig:remesh} for an illustration.

\begin{figure}
    \centering
    \includegraphics[width=0.9\textwidth]{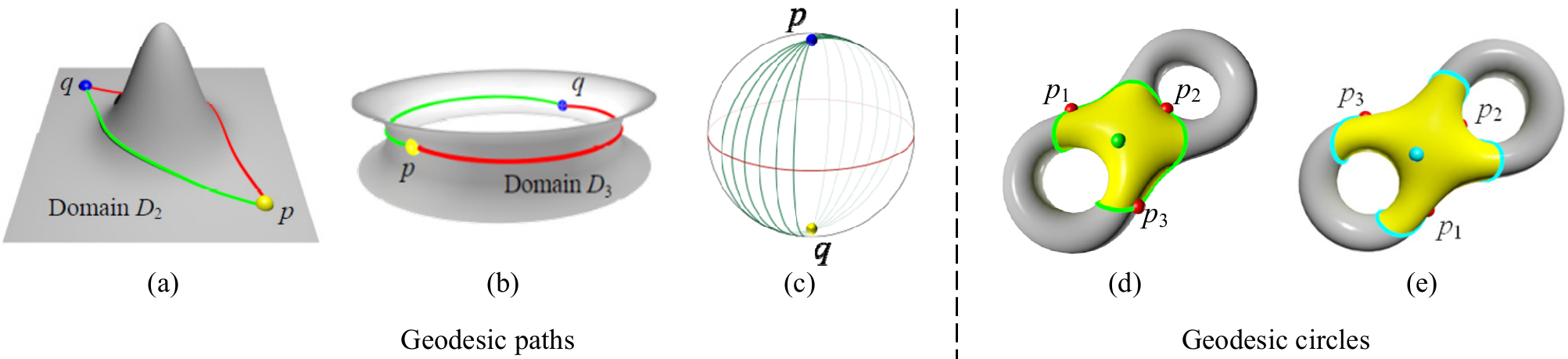}
    \caption{Geodesic paths and circles on curved manifolds. There are two different geodesic circles passing through three points $p_1$, $p_2$ and $p_3$ on $M$:
    one is in the front view (d) and the other is in the back view (e).}
    \label{fig:Euclidean-geodesic}
\end{figure}

\begin{figure}
    \centering
    \includegraphics[width=.4\textwidth]{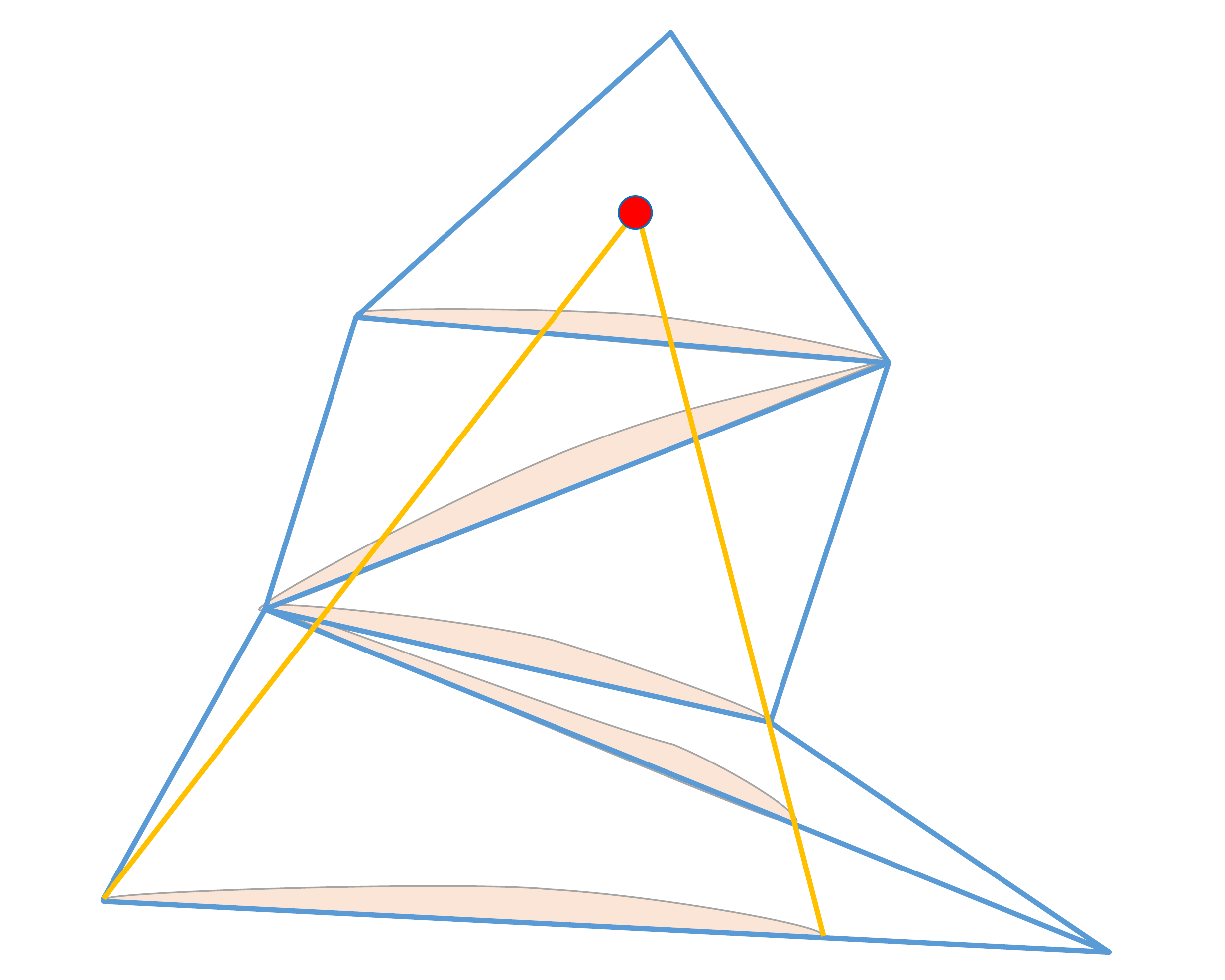}
    \caption{A visibility wedge (VW) on the bottom mesh edge.}
    \label{fig:visibility_wedge}
\end{figure}

\section{Theories}
\label{sec:theory}

Geodesic paths are locally shortest paths between any two points on the manifold. Due to the bending of non-zero curvatures on curved manifolds $M$, the distance field on $M$ characterized by geodesic distances/paths have quite different structures from that in Euclidean space $\mathbb{R}^n$, such as:
\begin{itemize}
\item Between any two non-duplicated points in $\mathbb{R}^n$, there is one and only one shortest path. However, between any two non-duplicated points on $M$, there may be one, two or infinite shortest paths; See Figure \ref{fig:Euclidean-geodesic} (a-c) for an example. Only in a smooth, simply connected 2-manifold with negative Gaussian curvature everywhere, the geodesic path betwteen any two points on $M$ is unique.
\item Given three points not lying on the same line in $\mathbb{R}^n$, there is a unique circle passing through them. However, given three points not lying on the same geodesic path on $M$, there may be no or more than one geodesic circles passing through them; See Figure \ref{fig:Euclidean-geodesic} (d-e) for an example.
\end{itemize}
Therefore, the combinatorial structure of geodesic Voronoi tessellations on $M$ are quite different from those in $\mathbb{R}^n$. Below we summarize the study of combinatorial structures on 2-manifold meshes $M_2$ in hierarchical way \cite{Yong2011Construction,Liu2013ipl,Liu2017CID}.

\subsection{Discrete geodesics}

Mitchell et al. \cite{Mitchell87} establish the discrete geodesic structure on $M_2$ (ref. Figure \ref{fig:visibility_wedge}):
\begin{itemize}
\item Inside every triangle in $M_2$, geodesic paths are straight line segments;
\item When crossing a mesh edge $e$, geodesic paths are straight lines if two adjacent faces of $e$ are unfolded in the same plane along $e$;
\item Starting from the triangle that contain the source point, a visibility wedge (VW) can be initialized and propagated across edges until all edges in $M_2$ are covered;
\item The mesh vertices lying on any geodesics (i.e., the apexes of any VWs) are called {\it pseudo-sources}, which can only be saddle vertices, i.e., the vertices in $M_2$ whose sum of surrounding angles is not smaller than $2\pi$.
\end{itemize}
The VW structure proposed in \cite{Mitchell87} can efficiently answer the single-source-all-destination discrete geodesic problem.

\begin{figure}
\centering
\subfigure[Isocontour structure]{\includegraphics[width=.35\textwidth]{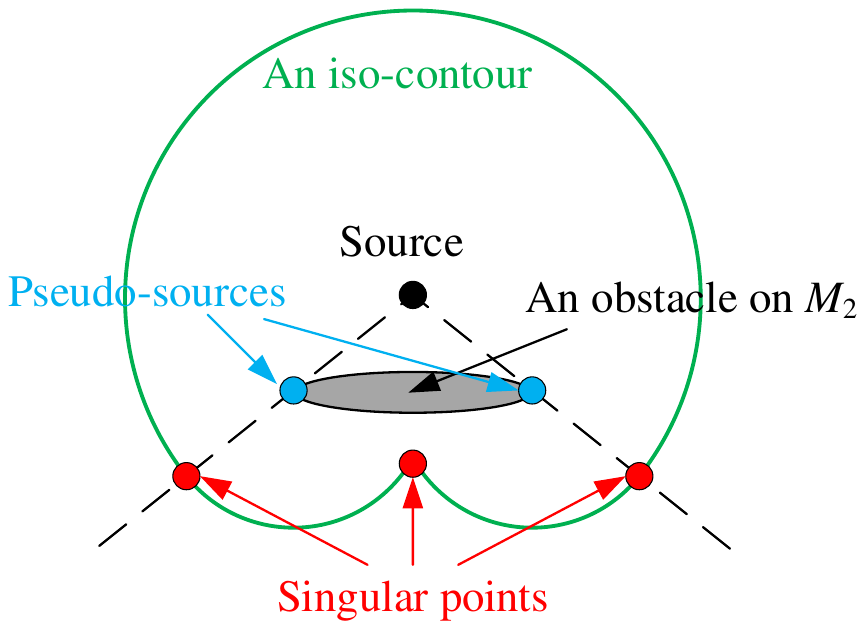}}\hspace{20pt}%
\subfigure[Real example]{\includegraphics[width=.32\textwidth]{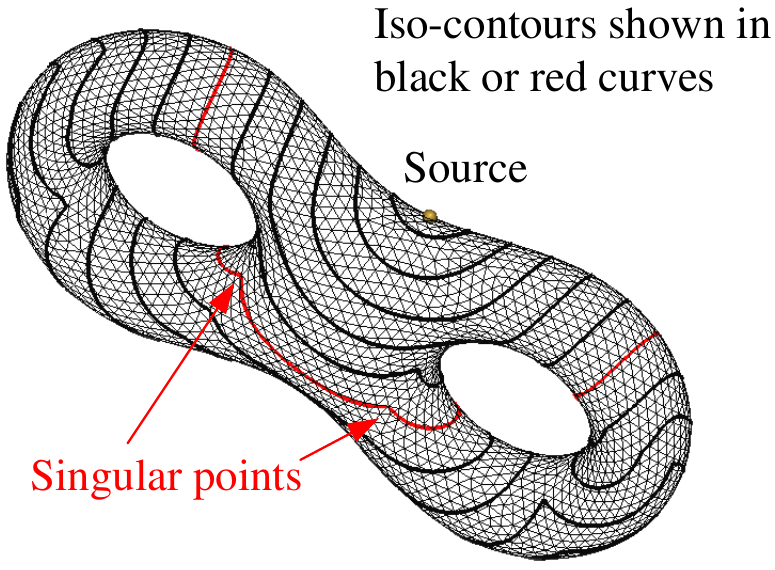}}%
\caption{Each iso-contour of the distance field on a closed $M_2$ consists of one or more closed curves and each closed curve consists of circular arc segments joined at singular points.
The obstacle in (a) can be a mountain shape with a sufficient height.}
\label{fig:isocontour}
\end{figure}

\begin{figure}
\centering
\subfigure[Bisector structure]{\includegraphics[width=.35\textwidth]{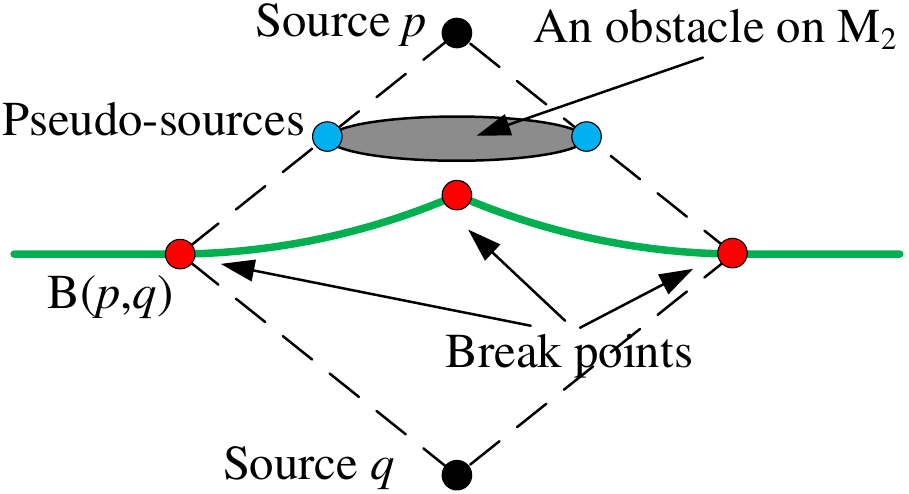}}\hspace{20pt}%
\subfigure[Real example]{\includegraphics[width=.28\textwidth]{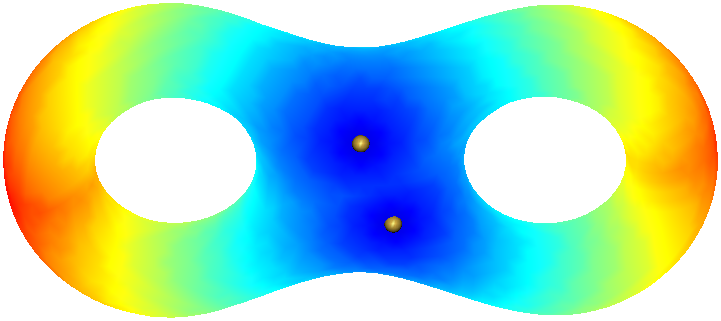}\hspace{10pt}\includegraphics[width=.28\textwidth]{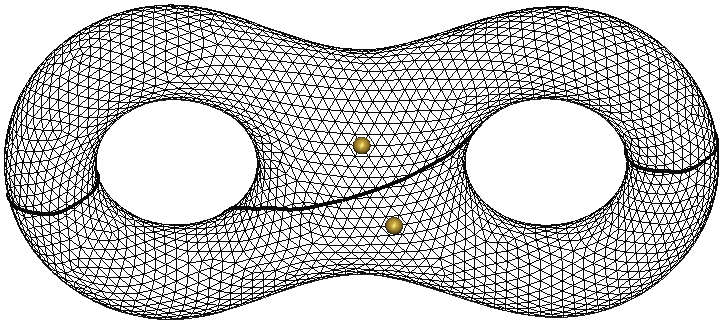}}%
\caption{(a) If all vertices do not have the same geodesic distance to a given pair of source points, their bisector consists of 1D curve segments, $C^0$ jointed at breakpoints. Between two breakpoints, the bisector portion is a line or hyperbolic segment.
(b) A real example: the geodesic distance field (left) and bisectors between two points (right).}
\label{fig:bisector}
\end{figure}

\subsection{Iso-contour structure}

Given one or more source points $P=\{p_i\}_{i=1}^K$, the geodesic distance is $d_g(x)=\min_i\{d_g(x,p_i),p_i\in P\}$, $\forall x\in M_2$.
An iso-contour (a.k.a. level set) of the distance field $d_g(x)$ is the trace of all points on $M_2$ that have the same distance value.
Iso-contours had drawn considerable attention in literature. On 2-manifold meshes $M_2$, their analytical structures are studied in \cite{Yong2011Construction} (ref. Figure \ref{fig:isocontour}):
\begin{itemize}
\item Due to the existance of pseudo-sources, each iso-contour of the distance field on a closed $M_2$ consists of one or more closed curves;
\item Each closed curve consists of circular arc segments joined at {\it singular} points, which are locations where the nearest pseudo-source is changing from one to another;
\item The number of closed curves in an isocontour depends on the indices of critical points of the distance field function, where a point $c\in M_2$ is a critical point of the distance field function $D$, if the partial derivatives of $D$ vanish at $c$. The index of a critical point $c$ is the number of negative eigenvalues of a Hessian matrix of $D$ at $c$.
\end{itemize}

\begin{figure}
\centering
\includegraphics[width=.52\textwidth]{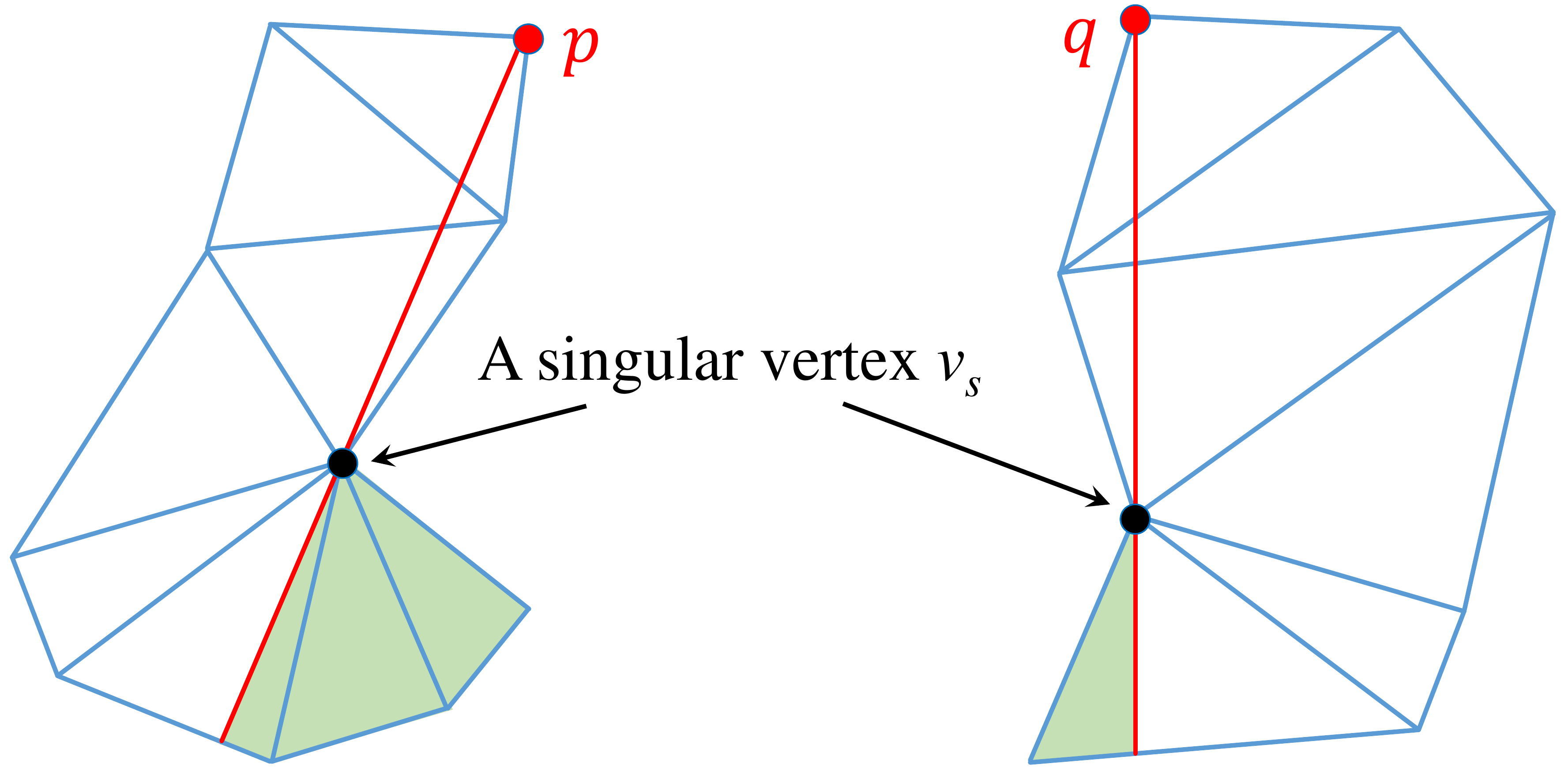}
    \caption{If a saddle vertex $v_s$ lies on the bisector of two points $p$ and $q$, i.e., $d_g(p,v_s)=d_g(q,v_s)$, this bisector contains a 2D regions (shaded area).}
    \label{fig:bisector-2D}
\end{figure}

\subsection{Bisector structure}

The bisector between any two points on $M_2$ is the trace of all points that have equal geodesic distance to these two points.
On 2-manifold meshes $M_2$, the analytical structure of bisectors are studied in \cite{Yong2011Construction} (ref. Figure \ref{fig:bisector}):
\begin{itemize}
\item The bisector between any two points $p$ and $q$ on $M_2$ may not be 1D. If a saddle vertex lies on the bisector, then this bisector contains a 2D region on $M_2$. See Figure \ref{fig:bisector-2D}. Another example can be found in \cite{Liu2015pami} (Figure 4);
\item Upon small perturbation on the vertices of $M_2$, we can assume all vertices do not have the same geodesic distance to a given pair of source points, and then their bisector consists of 1D curve segments;
\item If the bisector of two points is 1D, this bisector can have at most $g+1$ disjoint closed curves, where $g$ is the genus of $M_2$;
\item For each closed curve in a bisector, it can be decomposed at breakpoints, which are the locations where the nearest pseudo-source is changing along the bisector.
      The bisector is only $C^0$ at breakpoints. Between two adjacent breakpoints, the bisector portion can only be line or hyperbolic segment.
\end{itemize}

\begin{figure}
\centering
\includegraphics[width=.6\textwidth]{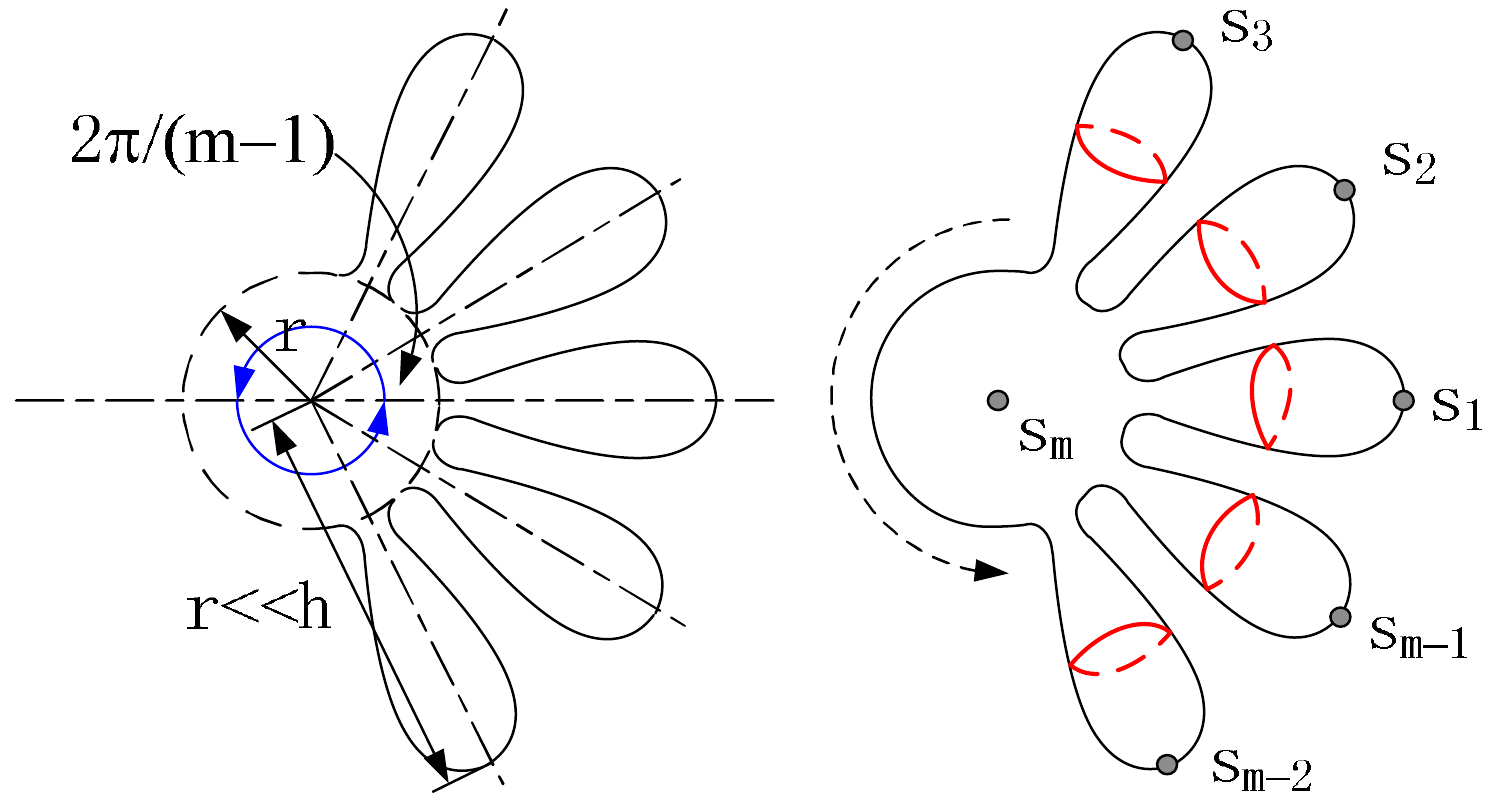}
    \caption{The geodesic Voronoi tessellation of the generator set $\{s_i\}_{i=1}^m$. The geodesic Voronoi cell of $s_m$ has $m-1$ closed Voronoi edges, each of which is a bisector between $s_m$ and $s_i$, $i=1,2,\cdots,m-1$.}
    \label{fig:octopus}
\end{figure}

\begin{figure}
\centering
\includegraphics[width=.35\textwidth]{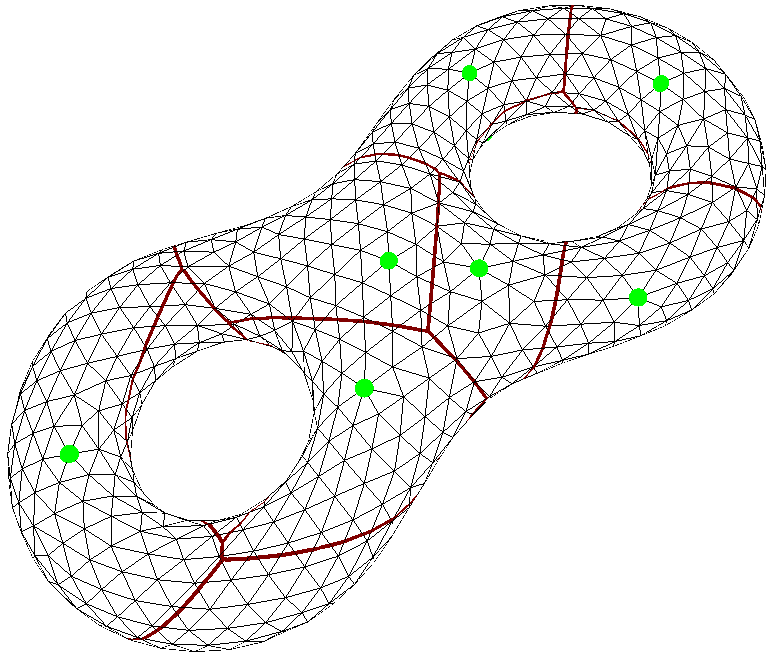}\hspace{20pt}%
\includegraphics[width=.35\textwidth]{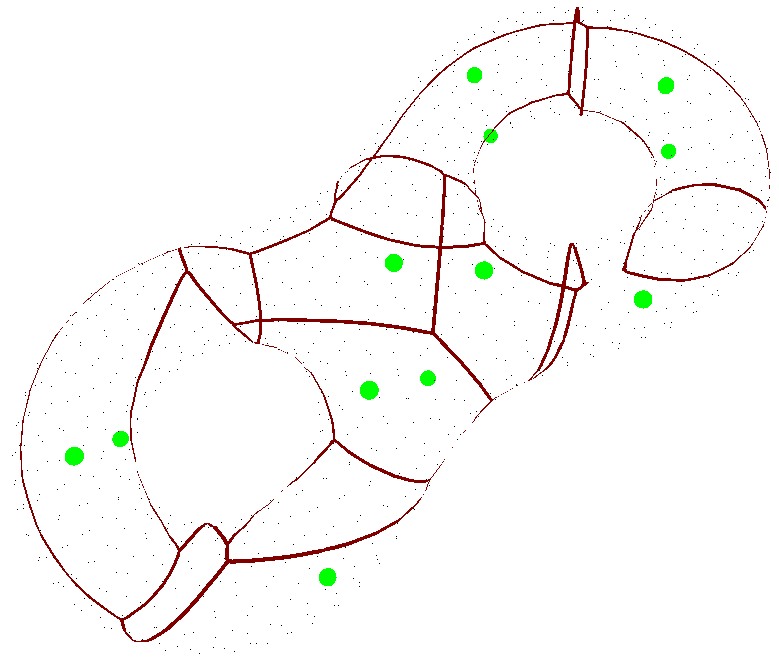}%
\caption{A real example of geodesic Voronoi tessellation of 12 point generators on a 2-manifold mesh with 2022 faces. Left is the geodesic Voronoi tessellation on the mesh. Right shows the boundary of each geodesic Voronoi cell with transparent surface rendering.}
\label{fig:vd}
\end{figure}

\begin{figure}
\centering
\includegraphics[height=.17\textwidth]{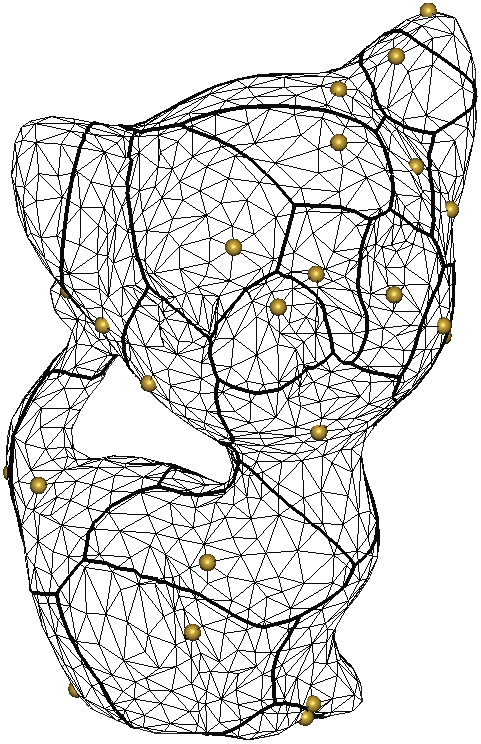}\hspace{6pt}
\includegraphics[height=.17\textwidth]{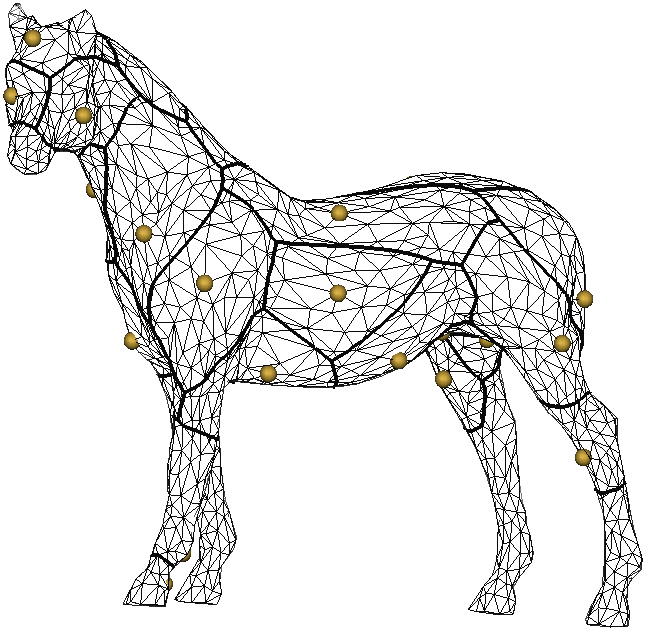}\hspace{6pt}
\includegraphics[height=.17\textwidth]{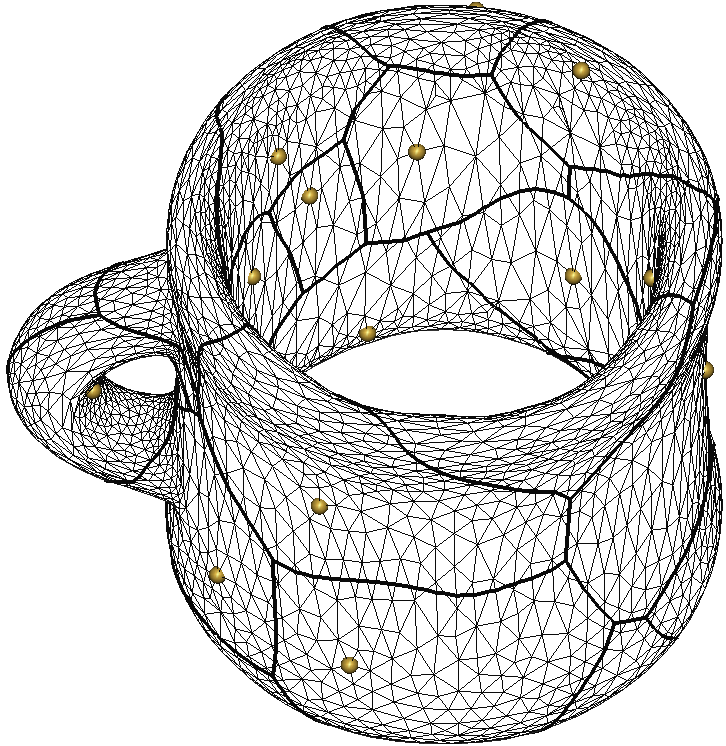}\hspace{12pt}
\includegraphics[height=.17\textwidth]{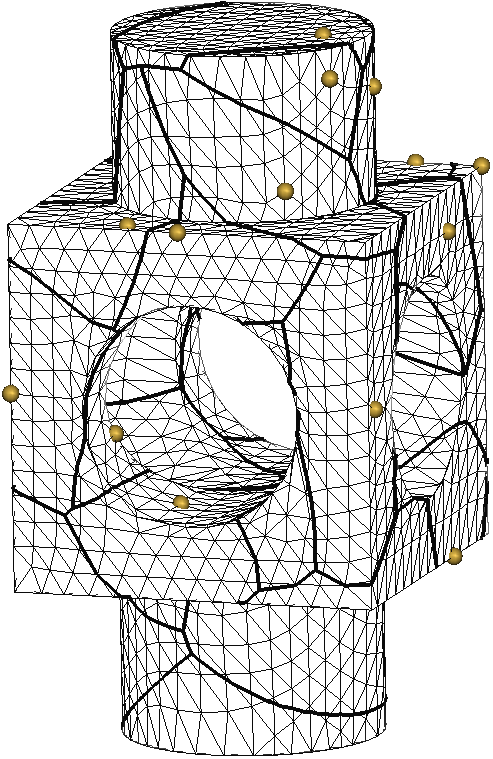}\hspace{6pt}
\includegraphics[height=.17\textwidth]{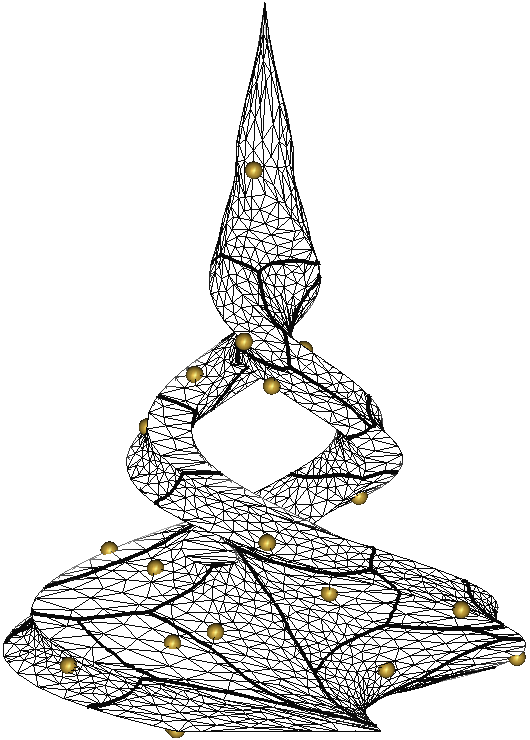}\hspace{6pt}
\includegraphics[height=.14\textwidth]{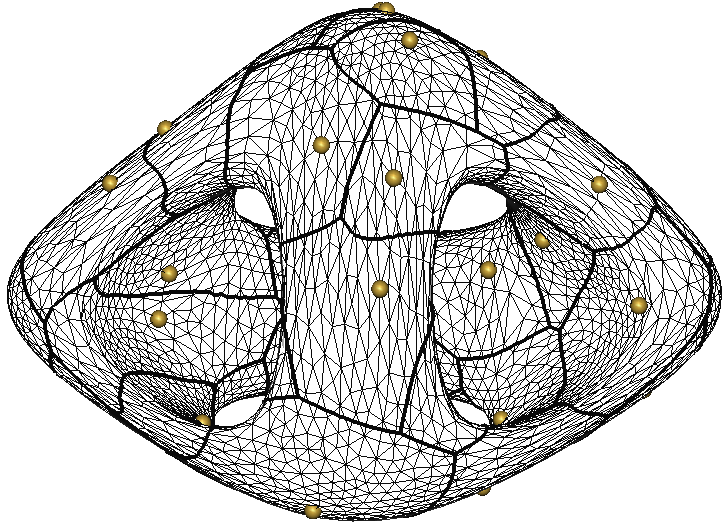}\hspace{6pt}
\caption{Some examples of geodesic Voronoi tessellations.}
\label{fig:more-gvt}
\end{figure}

\subsection{Geodesic Voronoi Tessellation}

Given a set of generators, the trimmed bisectors among them partition $M_2$ into geodesic Voronoi cells. Their analytical structures are studied in \cite{Yong2011Construction} (ref. Figures \ref{fig:octopus} and \ref{fig:vd}):
\begin{itemize}
\item Each geodesic Voronoi cell is connected, but may not be singly connected;
\item Each geodesic Voronoi cell is bounded by one or more closed curves called Voronoi edges. Each Voronoi edge consists of trimmed bisectors.
      Trimmed bisectors are joined at branch points that are locations on $M_2$ having the same geodesic distance to its three closest generators.
      One Voronoi edge does not have to contain a branch point.
\end{itemize}

It is well known that in $\mathbb{R}^2$, given a set of $n$ generators, there are at most $3n-6$ Voronoi edges and $2n-5$ branch points (also called {\it Voronoi vertices}) in Voronoi tessellation. The combinatorial structure of geodesic Voronoi tessellation on $M_2$ is studied in \cite{Liu2013ipl}:
\begin{itemize}
\item On a genus-$0$ $M_2$, the number of Voronoi vertices and Voronoi edges is $O(m)$, where $m$ is the number of generators;
\item The combinatorial complexity of geodesic Voronoi tessellation is defined to be the total number of breakpoints, Voronoi vertices, Voronoi edges and Voronoi cells.
      On a genus-$0$ $M_2$, the combinatorial complexity of geodesic Voronoi tessellation is $O(mk)$, where $k$ is the number of faces in $M_2$;
\item On a genus-$g$ $M_2$, the number of Voronoi vertices and Voronoi edges is $O(m+g)$, where $g$ is the genus of $M_2$;
\item If the set of generators is dense and the geodesic Voronoi tessellation satisfies the closed ball property \cite{Edelsbrunner1997}, the combinatorial complexity of geodesic Voronoi tessellation on a genus-$g$ $M_2$ is $O((n+g)k)$.
\end{itemize}

Some real examples of geodesic Voronoi tessellations are illustrated in Figure \ref{fig:more-gvt}.

\section{Algorithms}
\label{sec:algorithms}

Given the number $N$ of point generators on a $d$-manifold mesh, the GCVT can be computed by finding a tessellation that minimizes the GCVT energy in (\ref{eq:gcvt_energy}).
In this section, we present algorithms focusing on $d=2$. Some methods summarized in this section (e.g., the RCVT approximation method in Section \ref{subsubsec:rcvt}) can be extended to arbitrary $d\geq 2$ dimensions.

The energy in (\ref{eq:gcvt_energy}) can be minimized globally or locally. There are two existing global optimization methods for CVT: one is the Monte Carlo with minimization (MCM) framework \cite{tvcgLuSPW12} that only deals with CVT in $\mathbb{R}^2$ and the other is the manifold differential evolution (MDE) method \cite{Liu2016siga} that is general to deal with GCVT on $M_2$ ($\mathbb{R}^2$ is a special case of $M_2$). MCM is a heuristic method without theoretical guarantee, while MDE has a provable probabilistic convergence to the global optimum. In Section \ref{subsec:global-method}, we summarize the MDE global method.

Although the global method can achieve high-quality GCVT results and is insensitive to the initial position of generators, it is very time-consuming. Therefore, several fast, local optimization methods had proposed and we summarize two local methods in Section \ref{subsec:local-methods}.

\subsection{MDE Global Method}
\label{subsec:global-method}

MDE is a stochastic global optimization method, which extends the classic differential evolution \cite{Das2011} to the manifold setting.

\begin{figure}
    \centering
    \includegraphics[width=\textwidth]{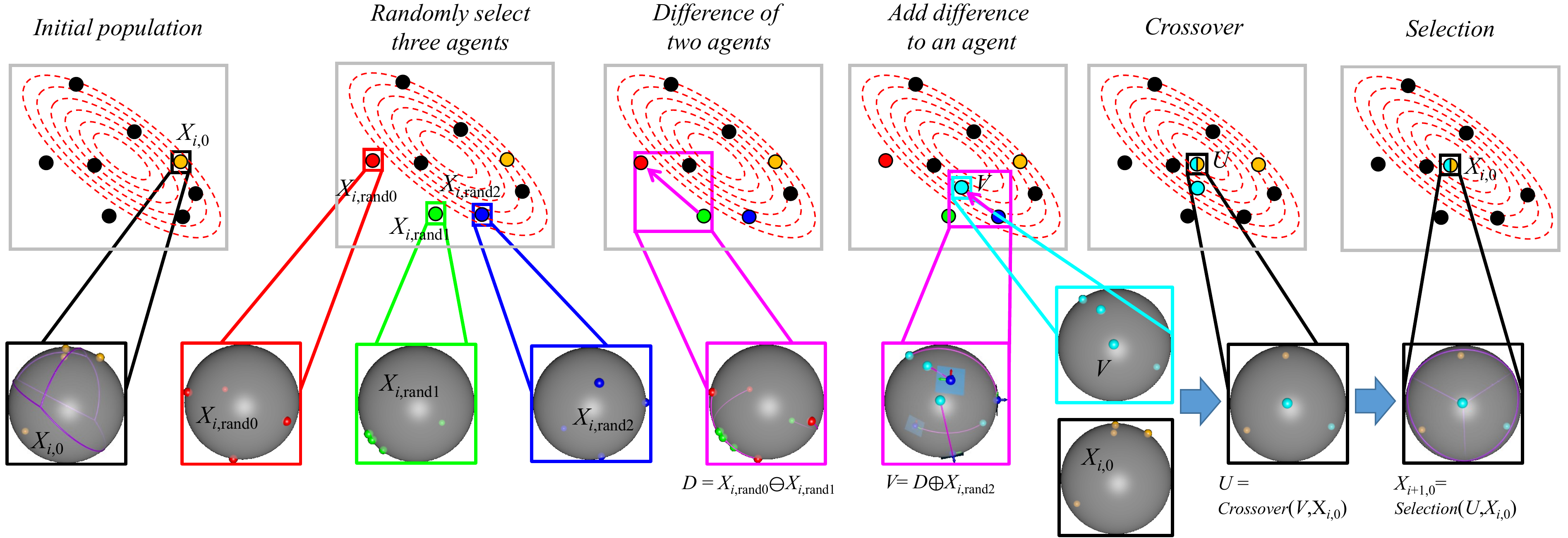}
    \caption{MDE pipeline.}
    \label{fig:MDE_illu}
\end{figure}

Classic differential evolution applies agents that have operations of addition, subtraction and scala multiplication, all defined in a vector space. To use these operations on a manifold, MDE assigned an order to the generators in an agent for encoding them into a vector representation such that different agents can be matched akin to matching vectors. The pipeline of MDE is illustrated in Figure \ref{fig:MDE_illu}. It initializes a group of agents by random sampling and improves the quality of the agents iteratively. There are three steps in each iteration of MDE, i.e. mutation, crossover and selection.

\emph{Vector representation and agent matching.} Given any two agents $G = \{ g_i \}_{i=1}^n$ and $G' = \{ g_i' \}_{i=1}^n$, MDE builds a complete bipartite graph $K_{n, n}$ whose vertices are $G \cup G'$ and edges weights are the geodesic distances between the pair of any two $g_i$ and $g_j'$. MDE finds a perfect matching in $K_{n, n}$ by solving the minimum-weight perfect matching problem using the Hungarian algorithm \cite{Korte2000Combinatorial}, which runs in $O(n^3)$ time. An order of generators can be induced by this matching, and thus the subscripts of generators can be corresponding by rearrangement. Let $X_{k, j} = \{ x_{k, j, i} \}_{i=1}^N, j = 1 \dots M$ be the $j$-th agent of $k$-th generation, where $N$ is the number of generators and $M$ is the number of agents.

\emph{Mutation operator.} MDE produces $M$ new \emph{mutative} agents by the mutation operator. The $j$-th mutative agent can be obtained by randomly selecting three agents and adding a scaled difference between two agents to another agent, i.e.
\begin{equation}
    V_{k, j} = X_{k, rand3} \oplus ( \lambda \otimes (X_{k, rand1} \ominus X_{k, rand2})),
\end{equation}
where $\ominus(y, x) : M \times M \rightarrow T_xM$ outputs a tangent vector at $x$ whose direction is the starting direction of the geodesic path from $x$ to $y$ and whose magnitude is length of the geodesic path, $\otimes$ is the scala multiplication in the tangent space and $\oplus(z, v) : M \times T(M) \rightarrow M$ obtains a point on the manifold by (1) parallel transporting $v$ to $z$ along the geodesic path from the point of $v$ to $z$, and denote the resulting vector as $v'$; (2) computing a geodesic path from $z$ with a initial direction $v'$ and the length is equal to $v'$, and the end point of this geodesic path is the output.

\emph{Crossover operator.} For each agent $X_{k, j}$, MDE produces a \emph{competitor} $ U_{k, j} = \{ u_{k, j, i} \}_{i=1}^N$ using the corresponding mutative agent $V_{k, j}$ by crossover operator. $u_{k, j, i}$ randomly uses the corresponding component of $X_{k, j}$ or $V_{k, j}$, i.e.,
\begin{equation}
    u_{k, j, i} =
\begin{cases}
v_{k, j, i} & rand_{(0,1)} < C_r\\
x_{k, j, i} & otherwise
\end{cases}
,
\end{equation}
where $C_r$ is \emph{crossover rate}.

\emph{Selection operator.} MDE selects a better agent from $X_{k, j}$ and $U_{k, j}$ and puts it into the new generation, i.e.,
\begin{equation}
    X_{k+1, j} =
\begin{cases}
X_{k, j} & \varepsilon(X_{k, j}) < \varepsilon(U_{k, j})\\
U_{k, j} & otherwise
\end{cases}
,
\end{equation}
where $\varepsilon$ is the GCVT energy of the agent computed by Equation~\ref{eq:gcvt_energy}.

The terminate condition of MDE is meeting one of the three conditions: (1) the iteration number exceeds the parameter specified by user; (2) the solution does not improve in successive several iterations; and (3) the GCVT energy reaches the prescribed value.

It was shown in \cite{Liu2016siga} that under some mild assumptions, the MDE solution converges to the global optimization with probability 1.

\begin{figure}
\centering
\subfigure[Regular mesh]{\includegraphics[width=.22\textwidth]{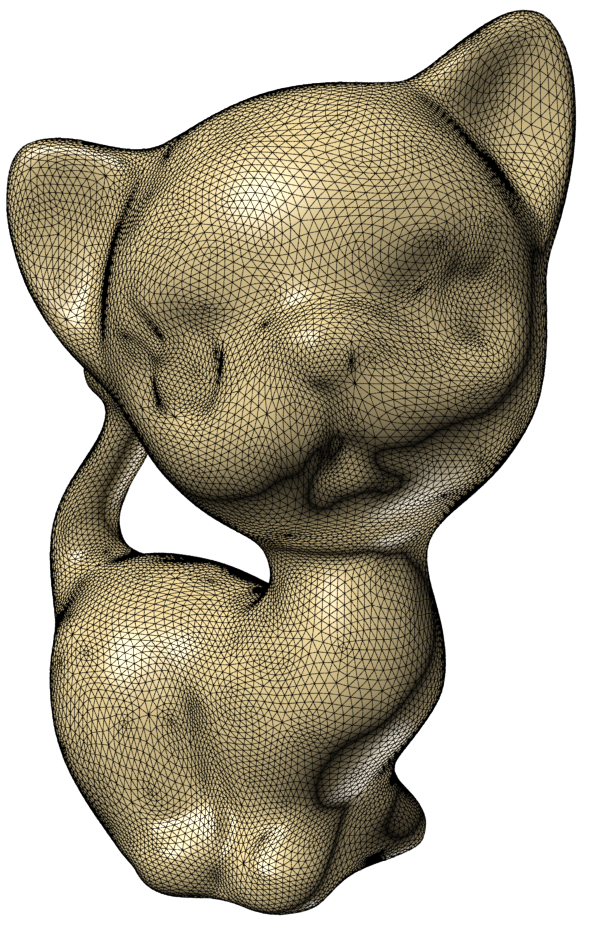}\hspace{6pt}\includegraphics[width=.22\textwidth]{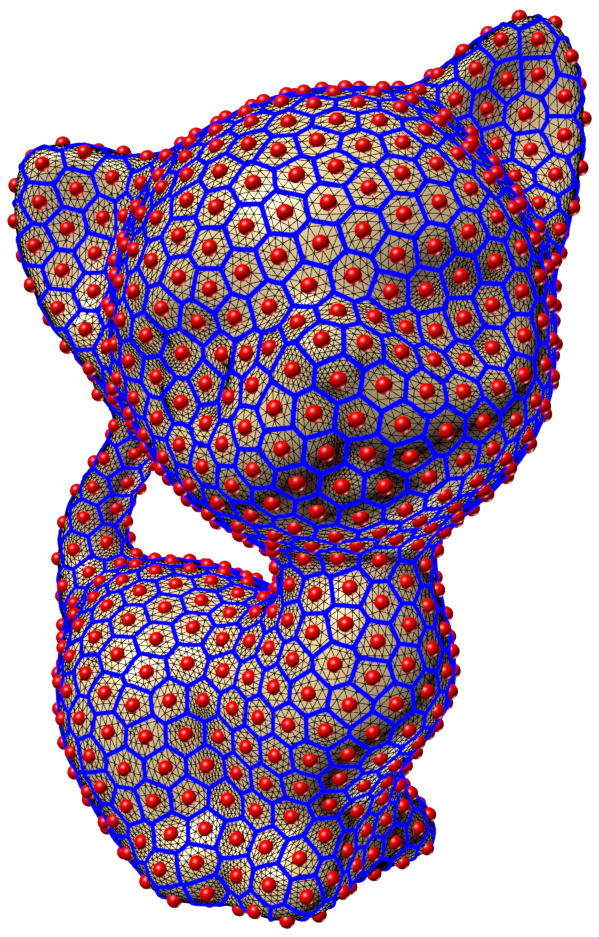}}\hspace{30pt}%
\subfigure[Highly irregular mesh]{\includegraphics[width=.22\textwidth]{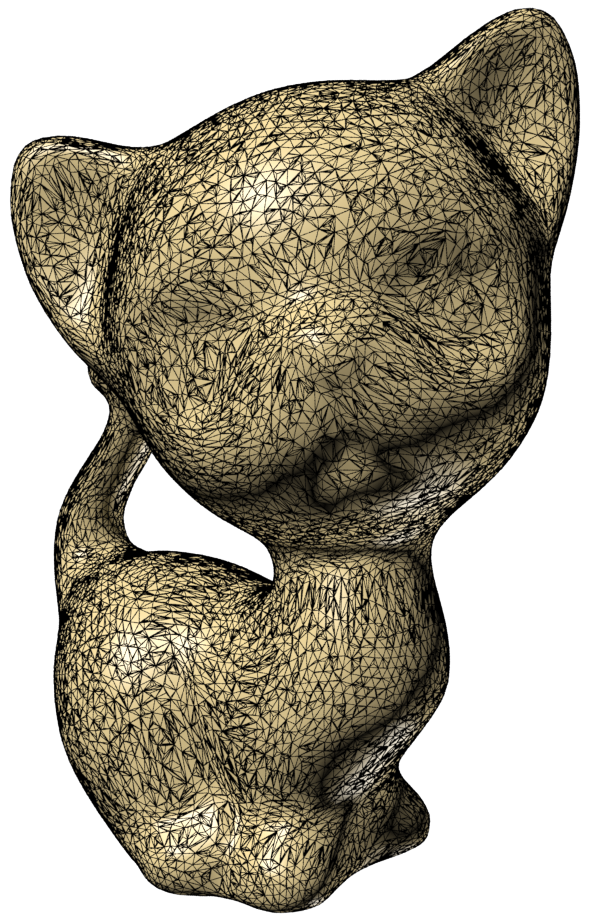}\hspace{6pt}\includegraphics[width=.22\textwidth]{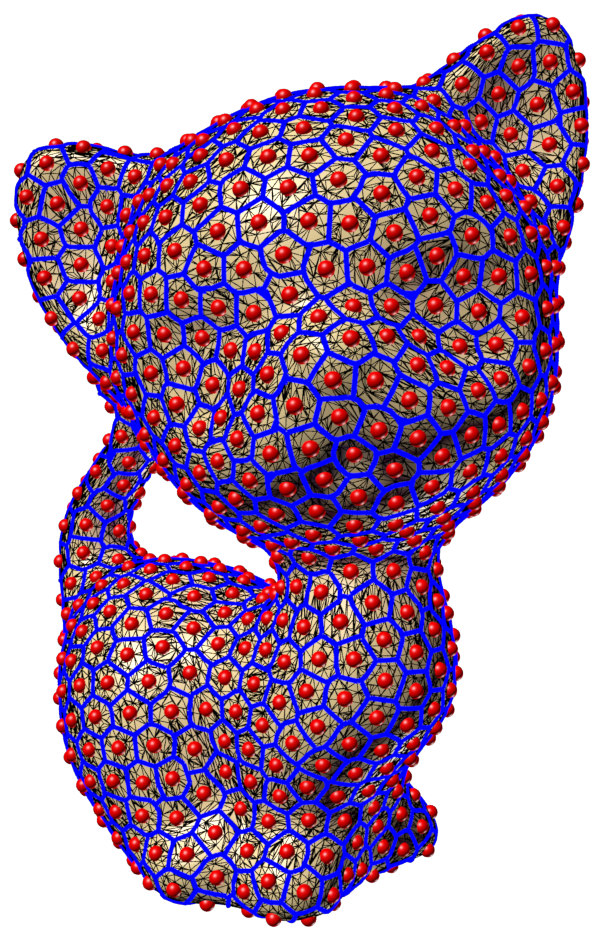}}
\caption{The regular mesh in (a) left and the irregular mesh in (b) left represent the same geometry. By applying the globally optimized MDE solution, the GCVTs on both meshes with the same number of point generators are the same.}
\label{fig:mde}
\end{figure}

The globally optimized MDE solution is insensitive to the underlying mesh quality. See Figure \ref{fig:mde} for an example.

\subsection{Two Local Methods}
\label{subsec:local-methods}

\subsubsection{Approximate nominal mass centroid}
\label{subsubsec:approx-centroid}

The Lloyd method \cite{Lloyd} is a classic algorithm that can be used to efficiently compute the cluster problem, including the construction of CVT and GCVT.
The Lloyd method locally minimizes the GCVT energy by iteratively moving the generators to the corresponding nominal mass centroids and updating the GVT of these generators.
I.e., in each iteration, there are two steps in Lloyd method: one is fixing the tessellation and the moving generators to the corresponding nominal mass centroids and the other is fixing the generators and updating the GVT. Both of the two steps reduce the GCVT energy \cite{Liu2017Intrinsic}, which ensures the convergence of the Lloyd algorithm.

The bottleneck of this Lloyd method lies on the computation of the nominal mass centroids, which requires to solve the optimization problem in (\ref{eq:nominal_mass_centroid}).
It is difficult or even impossible to solve this problem analytically and thus approximation has to be used.
Wang et al. \cite{WangYLXWGM015} propose an approximation method that computes the Riemannian center instead of solving the problem (\ref{eq:nominal_mass_centroid}).
Let $v_1,v_2,\cdots,v_m$ be the Voronoi vertices of a geodesic Voronoi cell on $M_2$. The Riemannian center is defined as the local minima of the following function:
\begin{equation}
E(x)=\sum_{i=1}^m d_g^2(x,v_i)
\end{equation}
Based on the properties studied in \cite{Pennec06,Rustamov10}, an iterative method utilizing the exponential map is proposed in \cite{WangYLXWGM015} to quickly find an approximation to the Riemannian center. Another approximation to the nominal mass centroid is proposed in \cite{Liu2017Intrinsic} that makes use of the landmark MDS (LMDS) \cite{SilvaT02} to quickly unfold a geodesic Voronoi cell into $\mathbb{R}^2$ in a way such that the total distance distortion defined by a graph embedding is minimized. Then the nominal mass centroid is approximated by the mass centroid (ref. Eq.(\ref{eq:mass_centroid})) of unfolded geodesic Voronoi cell in $\mathbb{R}^2$.

\subsubsection{RCVT approximation}
\label{subsubsec:rcvt}

The two local methods summarized in Section \ref{subsubsec:approx-centroid} make use of geodesic distance, which is time-consuming to compute for updating GVT in each Lloyd iteration.

Restricted centroidal Voronoi tessellation(RCVT) --- which utilizes Euclidean distance in embedded Euclidean space --- is proposed in \cite{Liu2016cvpr,Yi2018cvpr} as a fast approximation to GCVT.
Different from the two local methods in Section \ref{subsubsec:approx-centroid} that can only handle the 2-manifold meshes, the RCVT summarized in this section can deal with any $d$-dimensional triangulated meshes, $d\geq 2$.

Given a $d$-manifold $M$ embedded in $\mathbb{R}^n$, $d<n$, and a set of point generators $\{ g_i \}_{i=1}^k\in\mathbb{R}^n$ (not necessary on $M$), the \emph{restricted Voronoi region} corresponding to the point $g_i$ is defined as
\begin{equation}
\hat{V_i} = \{ x \in M\; \big|\; d_E(x, g_i) < d_E(x, g_j) \;\mbox{for}\, \ j=1,\dots,k,\,j \neq i \}.
\end{equation}
Elements of $\{ g_i \}_{i=1}^k$ are called \emph{generators}. The set of \emph{restricted Voronoi regions} $\{ \hat{V_i} \}_{i=1}^k$ forms a \emph{restricted Voronoi tessellation} of $M$.
The \emph{mass centroid} $m_i$ of $\hat{V_i}$ is defined by
\begin{equation}
\label{eq:restricted_mass_centroid}
m_i = \frac{\int_{x \in \hat{V_i}} xdx}{\int_{x \in \hat{V_i}} dx}.
\end{equation}
Similar to point generators $\{ g_i \}_{i=1}^k$, the mass centroid $m_i$ does not need to be on $M$.
The restricted Voronoi tessellation is called \emph{restricted centroidal Voronoi tessellation(RCVT)} if all the generators are mass centroids, i.e.
\begin{equation}
\label{eq:rcvt_condition}
g_i = m_i,\; i=1,\dots,k.
\end{equation}
Due to the following properties:
\begin{itemize}
\item Compared to GCVT, RCVT uses Euclidean distance and
\item Compared to CCVT, the mass centroids do not need to be on $M$,
\end{itemize}
RCVT behaves as a natural bridge between GCVT and CCVT.
Finally, RCVT is easy to compute using the Lloyd method with Euclidean distance.

\subsection{Algorithm Analysis}

In this section, we provide an analysis on the time complexity and space complexity of the above methods for constructing GCVT.

In MDE, the agent matching runs in $O(m^3)$ time and computing the exact geodesic between two points runs in $O(n^2\log n)$ time, where $m$ is the number of generators and $n$ is the number of vertices. Therefore, generating an agent takes $O(m^3 + n^2\log n)$ time and MDE runs in $O(N_pK(m^3 + n^2\log n))$, where $N_p$ is population size and $K$ is iteration number. It takes $O(N_pm)$ space for $N_p$ agents which have $N_pm$ generators in total.

IMSLIC \cite{Liu2017Intrinsic} and MSLIC \cite{Liu2016cvpr} construct GCVT and RCVT for an image using Lloyd method. It alternately constructs GVT and computes mass centroids. We analyze the complexities of the two steps separately. Fixing generators, IMSLIC can construct a GVT corresponding to the generators using Dijkstra's algorithm in $O(n\log n)$ time and $O(n)$ space. Adopting a label correcting method that maintains a bucket data structure can reduce the complexity and it only takes $O(n)$ time. Fixing generators, MSLIC can construct a RVT corresponding to the generators by going through all pixels in $O(n)$ time and $O(n)$ space. Fixing the tessellation, IMSLIC computes approximate nominal mass using LMDS in $O(n)$ time and $O(n)$ space and MSLIC computes mass centroids by going through all pixels in $O(n)$ time and $O(1)$ space. Therefore, we can obtain an approximate GCVT in $O(nK)$ time and $O(n)$ space and obtain an RCVT in $O(nK)$ time and $O(n)$ space for an image, where $K$ is the number of iteration.

\section{Conclusion}

Geodesic centroidal Voronoi tessellations (GCVTs) are intrinsic geometric structure inherent in manifolds. In this paper, we summarized our recent work on GCVTs on triangulated manifold.
We show our results from both theoretical and algorithmic aspects. Their applications in computer graphics and vision are also presented.
We hope that this paper can provide some insights for researchers to review the past developments and identify directions for future research on the study of intrinsic geometric structures in intelligent media data processing.

\bibliographystyle{unsrt}


\bibliography{references}

\end{document}